\documentstyle[pre,aps,twocolumn,epsfig,amsmath]{revtex}

\date{}

\title{Monte carlo simulations of parapatric speciation} 

\author{V.~Schw\"ammle$^{1,2}$, A. O. Sousa$^2$, S. M. de Oliveira$^1$}
\address{$^1$Instituto de F\'isica, Universidade Federal Fluminense,
Av. Litor\^anea, s/n, Boa Viagem, Niter\'oi 24210-340, RJ - Brazil}
\address{$^2$Institute for Computer Physics, University of Stuttgart,
Pfaffenwaldring 27, D-70569 Stuttgart, Germany}

\begin{document}

\maketitle

\begin{abstract}
Parapatric speciation is studied using an individual--based model with sexual 
reproduction. We combine the theory of mutation accumulation for biological ageing 
with an environmental selection pressure that varies according to the individuals 
geographical positions and phenotypic traits. Fluctuations and genetic diversity of 
large populations are crucial ingredients to model the features of evolutionary 
branching and are intrinsic properties of the model. 
Its implementation on a spatial lattice gives interesting insights into the population 
dynamics of speciation on a geographical landscape and the disruptive selection that leads 
to the divergence of phenotypes.
Our results suggest that assortative mating is not an obligatory ingredient to obtain 
speciation in large populations at low gene flow. 
\end{abstract}

\section{Introduction}

Several types of speciation are found in the literature, and the 
existence of some of them 
is still controversial. The two most discussed ones are the sympatric and the 
allopatric speciations.
The widely accepted mechanism of allopatric speciation is the appearance of a geographical 
barrier between two populations of the same species. Due to 
genetic drift and natural selection along several generations, these populations develop 
so many differences that they become reproductively isolated, that is, even if the barrier 
is removed the populations can no longer interbreed. In fact, speciation in allopatry 
is known to be a slow process~\cite{Mayr63}. The other form of speciation, the by far more 
complex sympatric speciation where there is no physical barrier to prevent gene flow, is 
supposed to be a fast process~\cite{Gavrilets2000}. Assortative mating 
(non--random mating) and  
competition for different niches seem to be its essential ingredients~\cite{Gavrilets2000,Lande82,Turner95,VanDoorn98,Higashi99,Kondrashov99,Dieckmann99,Luz2003,Arnegard2004}, 
although some authors claim that assortative mating alone is enough to produce reproductive  
isolation followed by sympatric speciation~\cite{Almeida2003}. On the other side, the model 
of~\cite{Porter2002} suggests 
that a small gene flux between different populations does not prevent them from speciation 
if the hybrids present a low viability. 

There have been great 
achievements to explain the processes of speciation in the last decade. The combination of 
laboratory experiments~\cite{Rice93}, measurements~\cite{Schliewen94,Schluter94,Via2001} 
and numerical models
~\cite{Gavrilets2000,Lande82,Turner95,VanDoorn98,Higashi99,Kondrashov99,Dieckmann99,Luz2003,Porter2002,Sa2001} 
gave enormous insights, especially into the theory of 
sympatric speciation and the processes driving it. However, less numerical research has  
been done focusing on parapatric 
speciation, a mixture of speciating in 
sympatry and in allopatry (for a review see~\cite{Gavrilets2004,Coyne2004}). The population 
occupies a spatially continuous habitat and adaptation evolves from a gradient, such as an 
increasing altitude or a continuous change of food resources 
\cite{Slatkin73,Endler73,Kirkpatrick97}, which 
may or not result in speciation \cite{Lande82,Sanderson89,Day2000}.

Here we modify the Penna model~\cite{Penna95,Moss99,Stauffer2001}, 
which is based on the mutation accumulation hypothesis for biological ageing,  
in order to include an environmental selection pressure that, besides acting according to 
individuals phenotypes, also varies according to their positions on a spacial lattice. 
Using this strategy we study under which conditions 
parapatric speciation happens and observe that it depends strongly on the fluctuations 
of the system, as already obtained in previous simulations of sympatric 
speciation~\cite{Luz2003,Sa2001}. The connection of the individual deaths with   
their phenotypic traits and lattice positions through a simple function is shown to produce 
a complex behavior of the whole population, that may or may not yield speciation.  

Our implementation of the sexual Penna model with a phenotypic trait on a spacial lattice 
is based on \cite{Sousa2004} and \cite{Luz2003}. We succeed in reproducing 
qualitatively the results of \cite{Gavrilets97}, although the effect of sexual 
selection in our model is shown to be so weak that it can be neglected. 

In the next section we explain our model, and in section 3 we present the results.  
In section 4 we discuss some relevant aspects of the model and section 5 contains the 
conclusions.  

%%%%%%%%%%%%%%%%%%%%%%%%%%%%%%%%%%%%%%%%%%%%%%%%%%%%%%%%%%%%%%%%%
%%%%%%%%%%%%%%%%%%%%%%%%%%%%%%%%%%%%%%%%%%%%%%%%%%%%%%%%%%%%%%%%%
\section{Model}
\label{sec:Penna}
\subsection{The Age Structured Part of the Genomes}

Genomes of diploid individuals are represented by two pairs of bit-strings, each string with 
32 bits. 
Individuals reproduce sexually and the strings of each pair are read in parallel (diploids). 
The first pair corresponds to the chronological genome of the Penna model and presents 
an age-structure. Each one of the 32 possible bit-positions of this pair represents a period 
in each individual's life, which means that each individual can live at most for 32 periods. 
Bits 1 correspond to a harmful recessive allele. If an individual carries two bits 1 at the 
same bit-position (homozygous), say position $i$, it means that the individual will start 
to suffer the effects of a genetic disease from its $i$-th period of life on. In dominant 
positions one bit set to one is enough to switch on a disease. At the beginning of the 
simulation we choose randomly 5 bit-positions to be the dominant ones. They are the same 
for all individuals and remain fixed during the whole simulation. 
At every iteration a new bit-position of all individual genomes is read; If the actual number 
of accumulated diseases of any individual reaches a threshold $T$, the individual dies.    

\begin{figure}[htb]
  \begin{center}
    \includegraphics[width=0.3\textwidth]{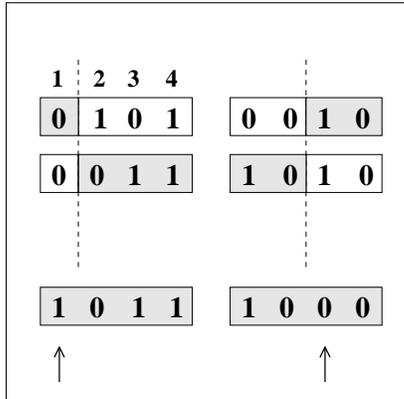}
\caption{Production of gametes. Each pair of  bit-strings is cut at a random
position and recombined. Mutations are then introduced at random positions in both 
parts (arrows). For the age-structured part only bad mutations are allowed, while for the 
non-structured part, both back and forward mutations can randomly occur.}
    \label{fig:crossover}
\end{center}
\end{figure}
At every iteration, females with age $\ge R$, the minimum reproductive age, search for a 
partner 
also with age $\ge R$ to breed, and produce offspring with a birth rate $b$. The 
offspring genome is constructed by crossing, recombination and mutation, as illustrated in 
Figure~\ref{fig:crossover}. The genome of the mother   
is cut at a random position and two random complementary pieces are joined to form a female 
gamete. Deleterious  mutations ($0\rightarrow1$) at random positions are then introduced, with 
a mutation rate $m$. The same process occurs with the father's genome and the union of the 
two gametes completes the offspring genome. In this part of the genome only deleterious 
mutations may appear, since they are 100 times more frequent than the backward mutations 
\cite{Pamilo87}. 
In this case, if the randomly chosen bit of the parent genome is already one, it remains one in 
the offspring genome (no mutation occurs). On the other hand, if the chosen bit is zero, 
it is set to one in the offspring genome.       

%%%%%%%%%%%%%%%%%%%%%%%%%%%%%%%%%%%%%%%%%%%%%%%%%%%%%%%%%%%%%%%%%%%%%%%%%%%
\subsection{Phenotypic Trait, Spatial Lattice and Ecology.}
\label{sec:mod}

The second pair of bit-strings of each genome is translated into some 
phenotypic characteristic of the individual \cite{Sa2001}, as, for instance, its size. 
This part also suffers crossing and recombination with mutations (Fig. \ref{fig:crossover}). 
However, for this part both good and bad mutations   
are allowed ($0\rightarrow1$ and $1\rightarrow0$), with a rate $m_p$. 
Moreover, 16 of the 32 bit-positions are dominant and 16 are recessive. The effective number 
of bits 1, taking into account the dominance, corresponds to a given phenotypic 
characteristic. This number, which we call the phenotype number $n$, is an integer between 
zero and 32. For example, we may consider that small values of $n$ correspond to small sized
individuals, while large values of $n$ denote big ones. 

The individuals are distributed on a two dimensional square lattice. They move 
at every iteration, with a rate $m_m$, to a randomly chosen less or equally populated nearest 
neighboring site. If all nearest neighbors sites are more populated than the current 
individual's site, the movement is not carried out. This 
strategy guarantees a fast and balanced distribution of individuals over the whole landscape. 
The reproductive females select their mating partners randomly from the reproductive males 
localized at the same or at a nearest neighbor site. Reproduction between different 
phenotypes is not forbidden. Offspring are distributed into empty 
nearest neighboring sites. If there is no empty site, the offspring is not produced. In 
this way the population size is controlled by the size of the lattice~\cite{Makowiec2001}, 
and there is no need to use the random killing Verhulst factor, present in the traditional 
version of the Penna model to avoid unlimited population growth.

\begin{figure}[htb]
  \begin{center}
    \includegraphics[angle=270,width=0.5\textwidth]{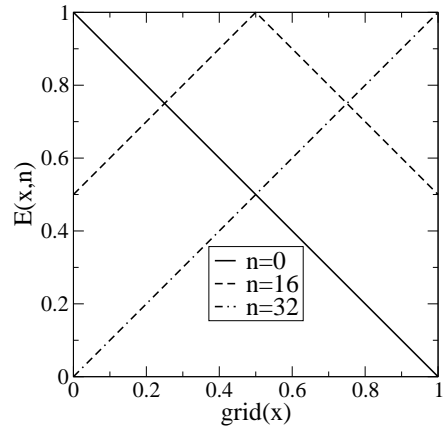}
\caption{Behavior of the ecological function (or probability to die according to 
         x-position and phenotype) $E(x,n)$. Individuals with high or low $n$ survive 
         better on opposite sides of the lattice whereas the ones with intermediate phenotype 
         numbers have a higher death probability everywhere.}
    \label{fig:EcolFunc}
\end{center}
\end{figure}

The interaction between phenotypic trait and geographical position on a square lattice 
of linear size $L$ is given by: 
\begin{equation}
    E(x,n) = S \cdot \left( 1 - \left| g(x)-\frac{n}{32} \right| \right),
\label{eq:ecology}
\end{equation}
which we call the ecological function. It gives the probability of an individual dying, 
at every iteration, depending on its x-position and phenotype number. 
The parameter $S$ is the strength of the interaction and varies between zero
and one. The larger the value of $S$ is, the stronger the selection pressure acting on the 
individuals. The coordinate 
function is given by $g(x) = \frac{x}{L-1}$, where the coordinate $x$ is an integer between 
zero and $L-1$.
For extreme phenotypes with $n=0$, the perfect region in which to live corresponds to 
$x=L-1$ where $E(L-1,0)=0$,  
while for extreme phenotypes with $n=32$ the perfect region corresponds to $x=0$. 
Individuals with intermediate phenotypes also live better at the extremes of the lattice,  
but are less fitted than those with extreme phenotypes living in the correct 
extreme of the lattice.  
Figure~\ref{fig:EcolFunc} illustrates the ecological function behavior for 
three different values of $n$. 

%%%%%%%%%%%%%%%%%%%%%%%%%%%%%%%%%%%%%%%%%%%%%%%%%%%%%%%%%%%%%%%%%%%%%%%%%%%
\section{Results}
\label{sec:results}

In this section we describe the relevant features of speciation found with 
our simulations, that is, we focus on the interaction between phenotypic trait and the lattice. 
For results of the traditional sexual Penna model with Verhulst factor or on a lattice we 
refer to~\cite{Moss99} and~\cite{Makowiec2001}, respectively. 

The fixed parameters that we adopt in the simulations are: 
 
\noindent i) Threshold number of genetic diseases $T=3$;

\noindent ii) Minimum reproductive age $R=8$; 

\noindent iii) Birth rate $b=4$;

\noindent iv) Rate of bad mutations in the chronological genome $m=1$;  

\noindent v) Number of dominant positions in the chronological genome $D=5$.

\noindent vi) Mutation rate of the phenotypic trait $m_p=0.15$ or $m_p=0.2$. 

\noindent vii) Number of dominant positions in the phenotypic trait $D_p=16$.
\medskip

The relevant parameters for speciation are the movement rate $m_m$, the lattice size $L$ 
and the strength $S$ of the environmental pressure. 

\begin{figure}[htb]
  \begin{center}
    \includegraphics[angle=270,width=0.5\textwidth]{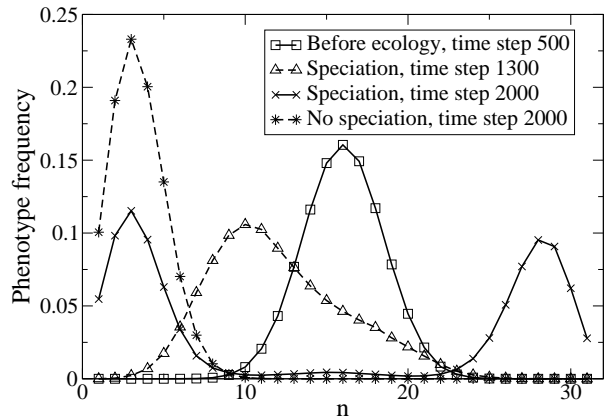}
\caption{The evolution of phenotypic frequency (given by the second pair of bit-strings) 
versus time. The same set of parameters may or not yield speciation for different random 
seeds. The central curve (solid line, squares) correspond to the distribution of phenotype 
numbers before switching on the ecological function (t = 1,000 time-steps). 
The final distribution when speciation 
occurs is given by the double-peaked distribution (solid line, crosses). The dashed line with 
stars corresponds to the final distribution for a case where speciation has not occurred; the 
dashed line with triangles shows the intermediary distribution in the course of 
speciation. The parameters: $S=0.24$, $m_m = 0.99$ and $m_p = 0.2$.}
    \label{fig:PhenFreq}
\end{center}
\end{figure}

We start the simulations with all the genomes randomly filled with zeroes and ones, and 
all individuals randomly distributed on the lattice. 
In order to reach a genetically stable initial population, we run the 
simulations without any ecological function for 1,000 iterations.
During this period the dynamics of the population is neither affected by the phenotype 
numbers nor by the lattice positions of the individuals. 
The initial distribution of the phenotype numbers is regulated solely by the 
mutations, and shows a Gaussian behavior (central curve of Fig. 
\ref{fig:PhenFreq}).
 
After these transient steps, the ecology is abruptly changed by setting the ecological  
function as an additional death probability. 
Disruptive selection driven by the ecology leads to a better survival of individuals 
with high and low phenotype numbers, depending on their current positions on the lattice. 
Three different situations, described below, can be observed, where the environmental pressure 
and the movement rate are the crucial parameters. 
\bigskip

i) At low selection pressures ($S$ small), and independently of the movement rate,  
the distribution of the phenotype numbers remains unaltered (Gaussian). The population 
decreases slightly at intermediate positions on the $x$--direction, but during the entire 
simulation individuals stay in contact over the whole lattice. Gene flow prevents 
disruptive selection from dividing the system into two sub-populations.
\bigskip

ii) For intermediate selection pressures and movement rates ($m_m \sim 1.0$), 
shortly after turning disruptive selection 
on, the system reaches an extremely dynamical state where fluctuations may or may not 
drive the system to divergence.
In the cases where speciation does not occur, the adaptation of the phenotypes  
on one of the lattice sides is faster and gene flow forces the individuals on the other 
side to adapt themselves to the opposite phenotype (dashed-line with stars in Fig. 
\ref{fig:PhenFreq}). 

When phenotypic adaption is balanced, the distribution of phenotype numbers bifurcates. 
Figure~\ref{fig:PhenFreq} shows that even in the case of speciation,   
the phenotypic distribution usually drifts away from symmetry before bifurcating, but 
the final and stable state corresponds to two populations 
with different phenotypes. We emphasize that during the 
speciation process the whole population stays in contact and gene flow can not be neglected 
as in allopatric speciation.

\begin{figure}[htb]
  \begin{center}
    \includegraphics[width=0.4\textwidth]{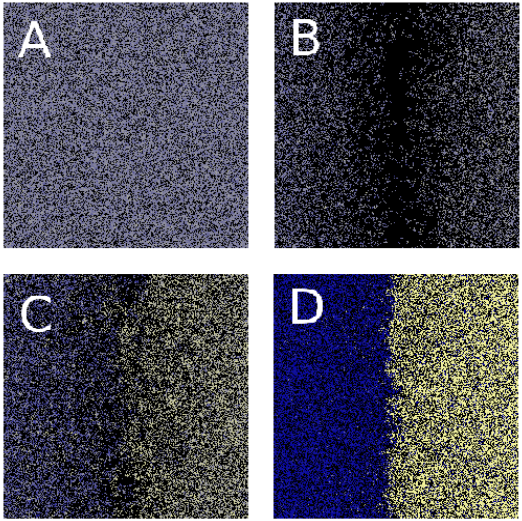}
\caption{Illustration of the phenotype distribution on a $500 \times 500$ lattice.
         The parameters are $m_m=0.9$, $S=0.25$ and $m_p = 0.1$.
         Black sites are empty. The colors indicate the average value of 
         the phenotype numbers (between yellow and blue). Without disruptive selection, 
         the initial population is homogeneously distributed over the whole lattice (A). 
         When the ecological function is turned on the population 
         is divided into two regions with weak contact (B). Selection prefers different 
         phenotypes with respect to the horizontal location of the individuals and adaptation 
         proceeds (C). 
         Fluctuations decide if the final result is speciation
         or if it is a single population with phenotypically similar individuals.  
         In case of speciation, two phenotypically different populations  
         can be easily distinguished, each one occupying one side of the lattice (D).}
    \label{fig:Fengrids}
\end{center}
\end{figure}

Figure~\ref{fig:Fengrids} shows the typical spacial 
distributions of the phenotypes at four different 
moments of the simulations. Initially, the population is homogeneously distributed over the 
whole lattice. As soon as the new ecology is turned on, almost all individuals occupying the 
central x-positions of the lattice die, and the population becomes temporarily divided 
into two similar groups, with weak contact between them. When the adaptation process of 
the extreme phenotypes starts, offspring with intermediate 
phenotype numbers continue to be produced. As the adaptation proceeds, competition with 
the more fitted extreme phenotypes makes the intermediate ones disappear.  
Finally, when speciation occurs, each half of the lattice becomes mostly occupied 
by one of the two extreme phenotypes, respectively. The number of iterations needed 
to reach the 
final distribution is about 5000, which corresponds to 625 generations. However, we would 
like to emphasize that we have run our simulations for up to 100.000 time-steps, to be sure 
we were obtaining stable distributions. 

\begin{figure}[htb]
  \begin{center}
    \includegraphics[width=0.4\textwidth]{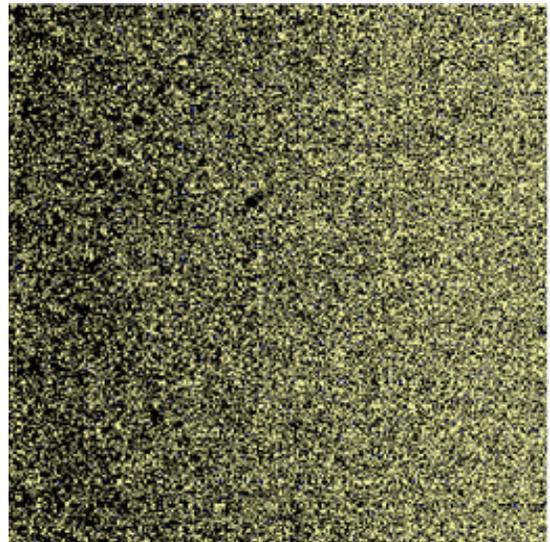}
\caption{Another random seed has driven the system to the case of no speciation. 
         One of the two sub-populations randomly dominates and finally occupies the 
         whole lattice. In this case, the worst region of the lattice for the winning 
extreme phenotype to survive (left side in this figure) remains less populated 
than the rest of the lattice, that is, the occupation is not uniform.}
    \label{fig:Fengrids2}
\end{center}
\end{figure}

The final result of a simulation where no speciation occurred, using the same parameters as in 
Figure~\ref{fig:Fengrids} but with another initial random seed, is illustrated in 
Figure~\ref{fig:Fengrids2}. In this case only one of the extreme 
phenotypes remains.  
\bigskip

iii) Low movement rates or very high selection pressures 
prevent speciation events. In both cases a  
great part of the population dies out at the time when the ecological function is set. 
Fluctuations dominate divergent adaptation and the initial 
Gaussian distribution of phenotypes moves to one of the extremes. 

It is important to say that for small population sizes fluctuations seem to always prevent 
speciation, independently of the movement rate: no speciation events have been obtained for 
lattice sizes smaller than $L=150$.

Concerning the selection pressure, it was also found by Doebeli \cite{Doebeli2003} that very 
high values of $S$ prevent speciation.

\begin{figure}[htb]
  \begin{center}
    \includegraphics[angle=270,width=0.5\textwidth]{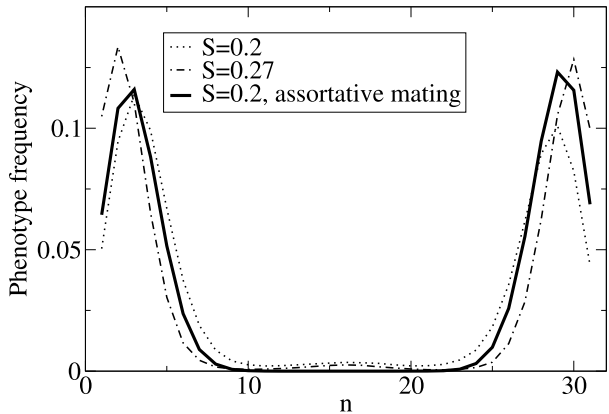}
\caption{Comparison of the final states of the phenotype distributions 
         for different values of the parameter $S$. The stronger the ecology is, the
         less frequent are the hybrids. Assortative mating leads to the non-existence 
         of hybrids, depicted by the thick curve where $d=10$. The movement rate $m_m=0.9$ 
         and the mutation rate related to the phenotype is $m_p=0.15$.} 
    \label{fig:PhenCompS}
\end{center}
\end{figure}

In order to study the effect of sexual selection in our simulations, we introduce  
the assortative mating strategy used in \cite{Gavrilets97} to prevent the mating of 
extreme phenotypes (prezygotic isolation). 
We measure the absolute difference of the phenotype numbers of both male and 
female, before mating. If the difference is larger than $d$, they can not reproduce. 
If there is no appropriate male among the nearest neighbors, no offspring 
is produced. 
Figure~\ref{fig:PhenCompS} shows the final phenotype distributions for 
different strengths $S$ of the ecological function, in cases where speciation 
occurred. We compare different results using random mating to one where assortative 
mating is used, with $d=10$. It can be seen that assortative mating completely prevents 
the production of hybrids with phenotype numbers around 16. Additionally, the occurrence 
of speciation is controlled by the parameter $d$, as in~\cite{Gavrilets97}. 
Very small values of $d$ ($d<8$) prevent speciation due to the lack of genetic diversity, 
which is an important ingredient for the distribution of phenotype numbers to bifurcate. 

\begin{figure}[htb]
  \begin{center}
    \includegraphics[angle=270,width=0.5\textwidth]{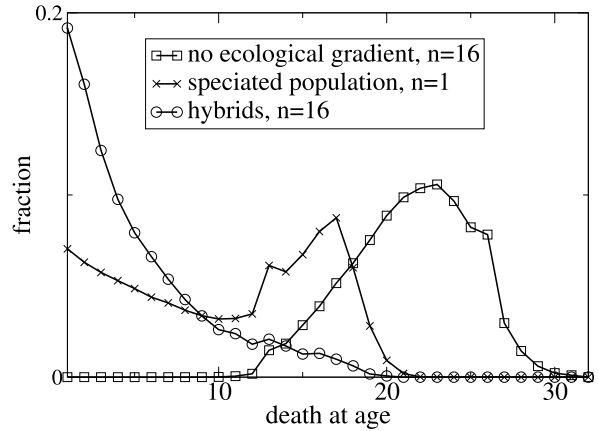}
\caption{Life span for different phenotype numbers. Most of the hybrids die at 
         low ages.}
    \label{fig:DeathAtAge}
\end{center}
\end{figure}

In Figure~\ref{fig:DeathAtAge} we show the histogram of the 
fraction of the population that 
dies at a given age, for different phenotype numbers. The majority of the hybrids die at 
low ages and do not generate offspring. These hybrids present low viability and thus 
characterize a speciation process \cite{Porter2002}.

\begin{figure}[htb]
  \begin{center}
    \includegraphics[angle=0,width=0.5\textwidth]{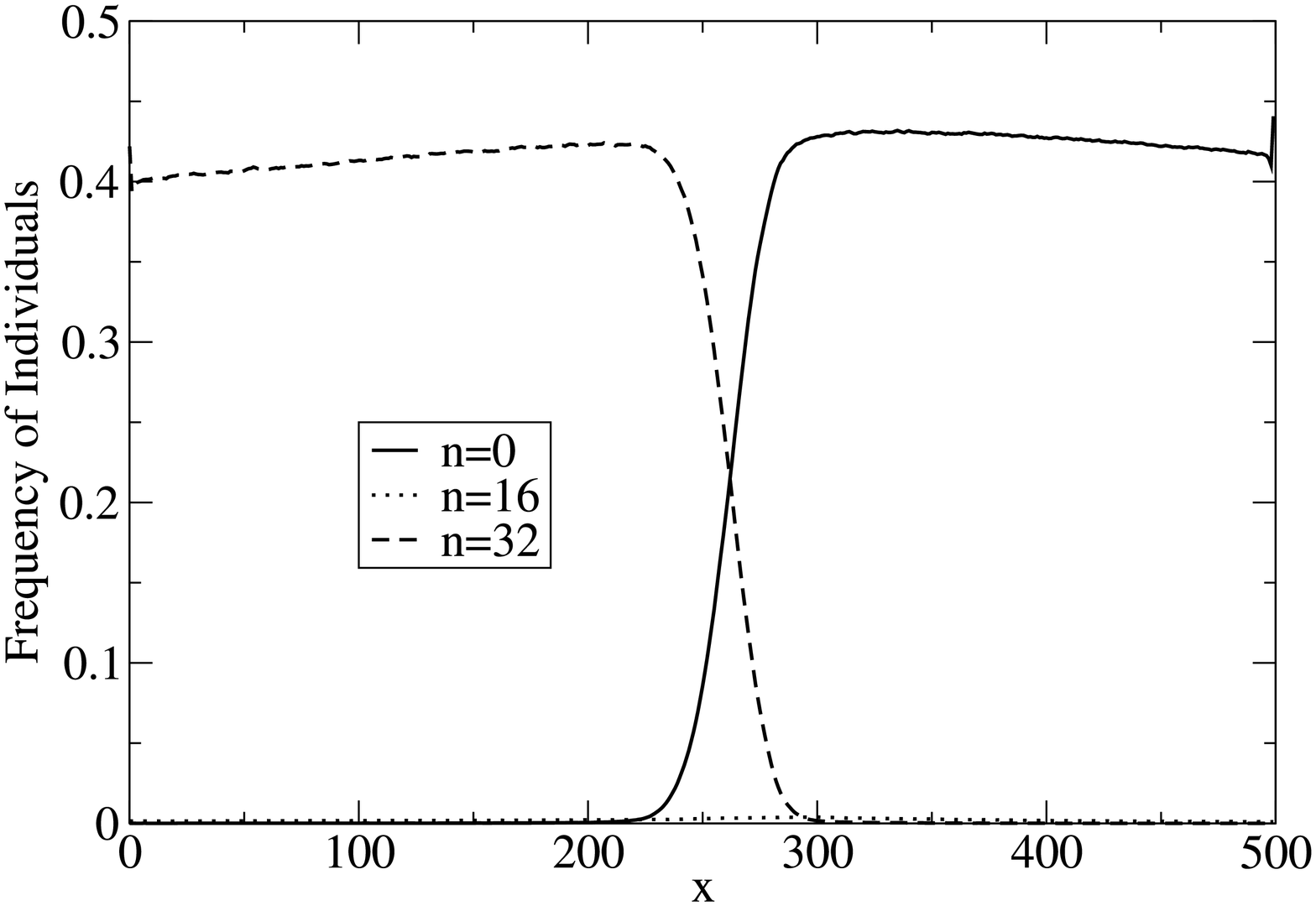}
\caption{Frequency of individuals with a given phenotype number for each position $x$ of the 
lattice, averaged over the last 10,000 time steps.}
    \label{fig:specx}
\end{center}
\end{figure}

A cline is defined as a gradient in a measurable character. Relative to the dispersal rate 
of a species, the slope of a cline between regions is indicative of the extent to which 
the inhabitants have differentiated. A steep cline means sharp differentiation while a gentle 
cline means indistinct divergence between areas \cite{Endler73}. In our case we choose the 
phenotype number, $n$, as the measurable character. Figure \ref{fig:specx} shows the fraction  
of individuals with $n=0$, $n=16$ and $n=32$ at each position $x$ of the lattice. A steep 
cline can be observed for the $n=0$ and $n=32$ populations, as well as the almost 
disappearance of the hybrids with $n=16$. 

\begin{figure}[htb]
  \begin{center}
    \includegraphics[angle=90,width=0.4\textwidth]{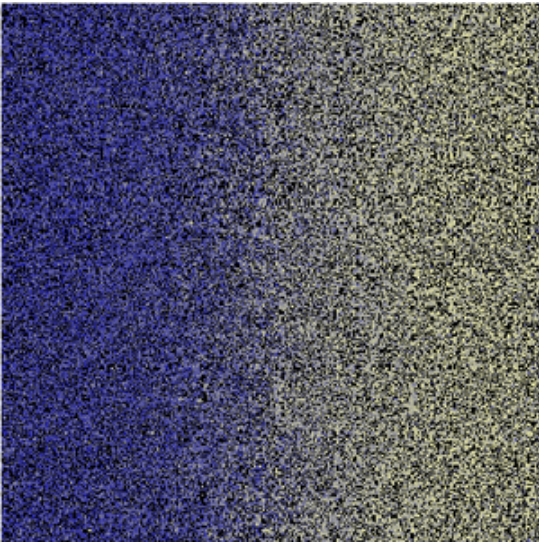}
\caption{Final state of the simulation without disruptive selection. No speciation occurs.}
    \label{fig:specy}
\end{center}
\end{figure}

\begin{figure}[htb]
  \begin{center}
    \includegraphics[angle=0,width=0.5\textwidth]{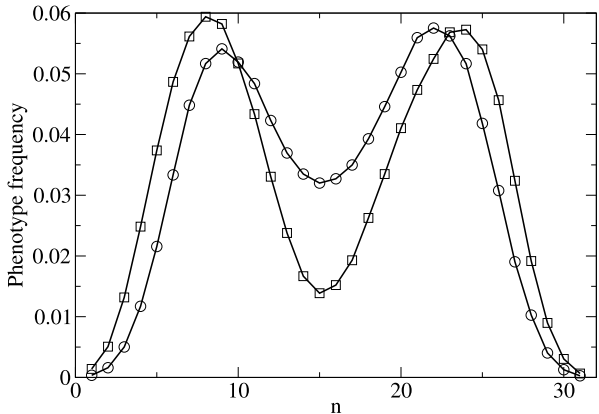}
\caption{Final phenotype distribution for simulations with the ecology function eq.(2). Even 
for a simulation with assortative mating gene flux prevents speciation.}
    \label{fig:specz}
\end{center}
\end{figure}

A common outcome in Nature consists of phenotypically distinguishable forms 
at geographic extreme regions and inter-grading hybrid forms in between. In our case 
disruptive selection due to the ecological function in eq.(1) prevents such a scenario.  
However, using the following ecological function: 
\begin{equation}
    E(x,n) = S \cdot \left| g(x)-\frac{n}{32} \right| ,
\label{eq:ecology2}
\end{equation}
hybrids are now   
favored and so do not disappear, that is, there is no speciation as shown in 
figures \ref{fig:specy} and \ref{fig:specz}. From fig. \ref{fig:specy} we can observe that 
the mean value of the phenotypes changes continuously with the geographic position $x$ and 
there is no sharp separation between the two extreme regions. 
It is important to say that in using eq.(2) speciation is not obtained even if assortative 
mating is included (fig. \ref{fig:specz}).

%%%%%%%%%%%%%%%%%%%%%%%%%%%%%%%%%%%%%%%%%%%%%%%%%%%%%%%%%%%%%%%%%%%%%%%%%%%
\section{Discussion}
\label{sec:disc}

As reported by Gavrilets~\cite{Gavrilets98} the dynamics of parapatric speciation is very 
fast (less than 1,000 generations) and is independent of the mutations rate $m_p$ of 
the phenotypic trait. We have made some simulations with smaller mutation rates and the time 
needed to reach a steady state did not increase for $m_p > 0.0001$. 

In our model the ecological 
function must be disruptive, i.e. individuals with intermediate phenotypes have to be 
discriminated. In a non--disruptive ecology individuals of all phenotypes can adapt to their 
local environments and hybrids evolve easily at intermediate x-positions. At the end 
of the simulations individuals of all phenotypes populate the lattice. 
If the selection against
individuals with intermediate phenotype numbers is not strong enough, only an unstable 
polymorphism appears: The two subpopulations coexist with large gene flux between them
until one of them completely dominates and uniformly occupies the whole lattice.

Different from other speciation models, ours allows fluctuations of all quantities, 
which hinders adaptation and the division of the system into two different phenotypic 
populations, even for intermediate values of the selection pressure.
This could explain the not so frequent occurrence of speciation in Nature, where many 
environmental factors act on the different population quantities, like the phenotypic 
distribution, and where fluctuations of these quantities are ubiquitous. Even if the 
conditions are optimal, speciation remains a 
statistical event (that is, for ten different initial random seeds, about five result 
in speciation and the other five result in an unimodal phenotypic distribution). Speciation is 
observed frequently for large lattices, where the phenotype 
distributions fluctuate less. 
Our results suggest that parapatric speciation occurs preferably in cases where a large 
population undergoes a sudden disruptive selection over large geographical distances 
compared to the range of individuals movements. 

We have studied the effect of assortative mating in our model, but the final results 
obtained were nearly the same as those using random mating, although 
the rule of ~\cite{Gavrilets97} increases the probability of speciation occurring.  
Even without assortative mating, only a very small number of hybrids is born (less than 
1\% of the total population)   
due to the small range of the mating region (only between nearest neighbors individuals). 
Moreover, Figure~\ref{fig:DeathAtAge} shows  
that these hybrids die mainly at low ages and do not produce offspring, which can be 
interpreted as a form of postzygotic reproductive isolation. 
In this way a small gene flow does not prevent speciation in this parapatric scenery, 
even without assortative mating. Models with small population 
sizes or mating over large geographical distances need assortative mating in order to obtain 
speciation~\cite{Gavrilets97,Doebeli2003}.

%%%%%%%%%%%%%%%%%%%%%%%%%%%%%%%%%%%%%%%%%%%%%%%%%%%%%%%%%%%%%%%%%%%%%%%%%%%
\section{Conclusions}
\label{sec:concl}

We present an individual-based model for parapatric speciation, where individuals with 
different phenotypes are distributed on a spatial lattice. Individuals may die due to genetic 
diseases or due to a competition for resources that depends on their phenotypes and on their  
geographical positions. Mating occurs only between next nearest neighbors. Surprisingly, 
even when 
considering random mating, fluctuations due to a disruptive selection may drive the system 
to speciation. On the other hand, under very strong disruptive selection, fluctuations 
prevent speciation to occur.

In fact, the importance of our approach is that it allows fluctuations in nearly all 
quantities. Physicists are very conscious about the importance of fluctuations in physical 
systems, mainly when they present a phase transition, which can be regarded as a process of 
bifurcation from a single phase (for instance, gas) into a state where two different phases 
coexist (liquid and vapor). The simplest, naive strategy to deal with such a phenomenon 
is the mean-field approach, in which the 
influences of the many units of the system over a particular one is replaced by an 
average influence or an ``average unity'', disregarding completely all possible fluctuations. 
However, this kind of treatment always gives wrong 
values for the critical exponents and sometimes signals the existence of a phase transition 
when it does not exist, since the fluctuations that would prevent the transition to occur are 
omitted \cite{Baxter82}. 

The speciation process is also a problem of bifurcation and mean-field approaches are
widespread (see for instance \cite{Kondrashov99,Dieckmann99}). Again, they certainly 
give the wrong speciation velocity (related to critical exponents) and may predict a 
speciation event when it does not exist. An example of mean-field approach as applied to
genetic evolution follows. Instead of considering the genetic features of each individual
separately, in a given generation, one considers the genetic frequency distribution of
the whole population and imposes some rule for its time evolution from one generation to
the other. Indeed, in this case, the genetic information of the whole population is
collapsed into a single ``average individual'', characterizing the above mentioned
mean-field approach. On the other hand, our model considers each individual separately,
and so does not ignore fluctuations. That is why speciation may or not appear depending
on the initial random seed, for the same set of parameters.
\bigskip

\section*{Acknowledgements}

V. Schw\"ammle is funded by the DAAD (Deutscher Akademischer Austauschdienst); S. Moss de 
Oliveira is partially supported by the Brazilian Agencies CNPq and FAPERJ.
We thank D. Stauffer, P.M.C. de Oliveira and J. S. S\'a Martins for a critical reading of the 
manuscript.

\bibliographystyle{revtex}
\bibliography{penna}
\end{document}